\documentclass[aps,prd,
,superscriptaddress,showpacs,footnote,preprint]{revtex4-1}

\usepackage[dvips]{color}
\usepackage{graphicx,subfigure}
\usepackage{amsmath}
\usepackage{amsfonts}
\usepackage{amssymb}

\begin{document}

\author{M. Orsaria\footnote{Home address: CONICET, Rivadavia 1917,
    1033 Buenos Aires, Argentina; Gravitation, Astrophysics and
    Cosmology Group Facultad de Ciencias Astron{\'o}micas y
    Geofisicas, Paseo del Bosque S/N (1900), Universidad Nacional de
    La Plata UNLP, La Plata, Argentina}}
\email{morsaria@rohan.sdsu.edu} \affiliation{Department of Physics,
  San Diego State University, 5500 Campanile Drive, San Diego, CA
  92182, USA}

\author{H. Rodrigues\footnote{Home address: Centro Federal de
    Educa\c{c}\~ao Tecnol\'ogica do Rio de Janeiro, Av
    Maracan$\tilde{a}$ 249, 20271-110, Rio de Janeiro, RJ, Brazil}}
\email{harg@sciences.sdsu.edu, harg@cefet-rj.br}
\affiliation{Department of Physics, San Diego State University, 5500
  Campanile Drive, San Diego, CA 92182, USA}

\author{F. Weber} \email{fweber@mail.sdsu.edu} \affiliation{Department
  of Physics, San Diego State University, 5500 Campanile Drive, San
  Diego, CA 92182, USA}

\author{G. A. Contrera} \email{guscontrera@gmail.com}
\affiliation{CONICET, Rivadavia 1917, 1033 Buenos Aires; Gravitation,
  Astrophysics and Cosmology Group, Facultad de Ciencias
  Astron{\'o}micas y Geofisicas, Paseo del Bosque S/N (1900),
  Universidad Nacional de La Plata UNLP, La Plata, Argentina}

\title{Quark-Hybrid Matter in the Cores of Massive Neutron Stars}

\begin{abstract}  
  Using a nonlocal extension of the SU(3) Nambu-Jona Lasinio model,
  which reproduces several of the key features of Quantum
  Chromodynamics, we show that mixed phases of deconfined quarks and
  confined hadrons (quark-hybrid matter) may exist in the cores of
  neutron stars as massive as around $2.1\, M_\odot$. The radii of
  these objects are found to be in the canonical range of $\sim 12-
  13$~km. According to our study, the transition to pure quark matter
  does not occur in stable neutron stars, but is shifted to neutron
  stars which are unstable against radial oscillations. The
  implications of our study for the recently discovered, massive
  neutron star PSR J1614--2230, whose gravitational mass is $1.97\pm
  0.04\, M_{\odot}$, are that this neutron star may contain an
  extended region of quark-hybrid matter at it center, but no pure
  quark matter. 
\end{abstract}

\date{\today}



\maketitle

\noindent {\it Introduction --} Quantum Chromodynamics (QCD) has the
properties of asymptotic freedom and confinement. The former implies
that in the high-momentum transfer regime, the quarks behave
essentially as free particles, i.e., the interaction between two
quarks due to gluon exchange is very weak. This regime can therefore
be treated using perturbation theory, where the quark-gluon coupling
constant serves as an expansion parameter.  For this momentum range,
the dispersion processes can be calculated very accurately. By
contrast, at low-momentum transfers ($\lesssim 1$~GeV) QCD becomes
highly nonlinear, which prevents the use of perturbative methods.  One
of the renowned effective models that serves as a suitable
approximation to QCD in the low-energy regime is the quark version of
the Nambu Jona-Lasinio (NJL) model \citep{nambu61:a,
  Vogl:1991,buballa04:a}. In this model chiral symmetry constraints
are taken into account via effective interactions between quarks,
through local four-point vertex interactions. The drawbacks of using
local interactions are that the model must be regularized to avoid
divergences in the loop integrals, and that the model is
non-confining. The absence of confinement is essentially related to the
fact that the dynamically generated constituent quark masses are
momentum independent. Since the 1990's, there have been investigations
proposing nonlocal interactions to solve these problems
\citep{Ripka97}.  One interesting suggestion arises from the
relationship between the NJL model and the model of one-gluon exchange
where an effective gluon propagator is used to generate effective
interactions between quarks. This provides a natural way to introduce
a nonlocality in the quark-quark interaction, which can be
characterized by a model-dependent form factor, $g(p)$
\citep{Blaschke94}.

In this paper we analyze the global structure and composition of
massive neutron stars in the framework of an extended version of the
nonlocal SU(3) NJL model. Of particular interest is the question as to
whether or not massive neutron stars, such as the recently discovered
pulsar PSR J1614--2230 whose gravitational mass was found to be
$1.97\pm 0.04\, M_{\odot}$ \cite{Demorest2010}, may contain stellar
cores made of deconfined quark matter \cite{Kim,Klahn,Masuda2012,Lenzi2012}.  Tendentially, one could
argue that quark deconfinement may not occur in the cores of such
high-mass neutron stars, since the underlying nuclear equation of
state must be extremely stiff in order to support such high mass
neutron stars. As will be shown in this paper, arguments along these
lines appear premature.

Our study is based on the latest set of model parameters of the
nonlocal SU(3) NJL model. The parameter fit is performed for a
phenomenological value of the strange quark mass of $m_s=140.7$
MeV. Details of the NJL formalism at zero chemical potential and
finite temperature can be found in
\cite{Scarpettini2004,Contrera2008,Contrera2010}.

Over the last two decades, several authors have started to take into
account the effect of the quark vector interaction in effective chiral
models like NJL\cite{Buballa:1996,Hanause:2000,Kitazawa:2002,%
    Klahn, Carignano:2010, Masuda2012, Lenzi2012}. It is known that
the repulsive character of the vector coupling in these models affects
the quark-hadron phase transition and moves the chiral restoration to
a larger value of the quark chemical potential \cite{Kitazawa:2002}. Thus,
if the quark deconfined transition in the cores of neutron stars is
modeled by a NJL-like model, it is expected that the vector coupling
contribution modifies the nuclear equation of state (EoS) and hence
the mass-radius relationship of neutron stars. Most of the
  NJL studies of neutron stars are treating the interactions among
  quarks in terms of local fermion-fermion couplings and/or impose the
  condition of local electric charge neutrality on the stellar matter
  \cite{Klahn,Masuda2012, Lenzi2012}.  In this paper, we are using a
  generalized version of the NJL model where the interactions are
  nonlocal and momentum-dependent, and the condition of local charge
  neutrality is replaced with the more relaxed condition of global
  charge conservation, as neutron stars are characterized by two
  rather than one conserved charge \cite{glendenning92:a}. As a
  consequence, the pressure in the mixed quark-hadron phase varies
  with density and is, therefore, not a priori excluded from neutron
  stars.

To include the vector interaction in the NJL model we follow
\cite{Weise2011}.  However, we shall consider three different vector
fields, one for each quark flavor, instead of a single vector field
for all quarks.

\vskip 0.5cm
\noindent {\it Description of quark matter phase in the framework of
  the nonlocal SU(3) NJL model --} We start from the Euclidean
effective action associated with the nonlocal SU(3) quark model,
\begin{eqnarray}
S_E &=& \int d^4x \ \{ \bar \psi (x) \left[ -i \gamma_\mu
\partial_\mu + \hat m \right] \psi(x)\nonumber \\ 
& & - \frac{G_s}{2} \left[
j_a^S(x) \ j_a^S(x) + j_a^P(x) \ j_a^P(x) \right]\nonumber \\ 
& & - \frac{H}{4} \ T_{abc} \left[
j_a^S(x) j_b^S(x) j_c^S(x) - 3\ j_a^S(x) j_b^P(x) j_c^P(x) \right]\nonumber \\ 
& & - \frac{G_{V}}{2} j_{V,f}^\mu(x) j_{V,f}^\mu(x) , \label{se}
\end{eqnarray}
where $\psi$ is a chiral $U(3)$ vector that includes the light quark
fields, $\psi \equiv (u\; d\; s)^T$, and $\hat m = {\rm diag}(m_u,
m_d, m_s)$ stands for the current quark mass matrix. For simplicity we
consider the isospin symmetry limit, in which $m_u = m_d=\bar m$. The
fermion kinetic term includes the convariant derivative
$D_\mu\equiv \partial_\mu - iA_\mu$, where $A_\mu$ are color gauge
fields, and the operator $\gamma_\mu\partial_\mu$ in Euclidean space
is defined as $\vec \gamma \cdot \vec \nabla +
\gamma_4\frac{\partial}{\partial \tau}$, with
$\gamma_4=i\gamma_0$. The currents $j_a^{S,P}(x)$ and
$j_{V,f}^{\mu}(x)$ are given by
\begin{align}
  j_{a}^S(x) & =\int d^{4}z\ \widetilde{g}(z)\ \bar{\psi}\left(
    x+\frac{z}{2}\right)
  \ \lambda_{a}\ \psi\left(  x-\frac{z}{2}\right)  \ ,\nonumber\\
  j_{a}^P(x) & =\int d^{4}z\ \widetilde{g}(z)\ \bar{\psi}\left(
    x+\frac{z}{2}\right)
  \ i \ \gamma_5 \lambda_{a} \ \psi\left(  x-\frac{z}{2}\right)\ ,\nonumber\\
  j_{V,f}^{\mu}(x) & =\int d^{4}z\ \widetilde{g}(z)\
  \bar{\psi}_f\left( x+\frac{z}{2}\right) \ \gamma^{\mu}\ \psi_f\left(
    x-\frac{z}{2}\right), \label{currents}
\end{align}
where $\widetilde{g}(z)$ is a form factor responsible for the
non-local character of the interaction, and the matrices $\lambda_a$,
with $a=0,..,8$, are the usual eight Gell-Mann $3\times 3$ matrices --
generators of SU(3) -- plus $\lambda_0=\sqrt{2/3}\,
\mathbb{I}_{3\times 3}$. Finally, the constants $T_{abc}$ in the
t'Hooft term accounting for flavor-mixing are defined by
\begin{equation}
T_{abc} = \frac{1}{3!} \ \epsilon_{ijk} \ \epsilon_{mnl} \
\left(\lambda_a\right)_{im} \left(\lambda_b\right)_{jn}
\left(\lambda_c\right)_{kl}\;.
\end{equation}
After standard bosonization of Eq. (\ref{se}), the integrals over the
quark fields can be performed in the framework of the Euclidean
four-momentum formalism. The grand canonical potential in the
mean-field approximation at zero temperature, including the vector
coupling, is then given by
\begin{eqnarray}
  && \Omega^{NL} ( T=0,\mu_f) =  -\,\frac{N_c}{\pi^3}\,\sum_{f=u,d,s}\, \int^{\infty}_{0}
  \,dp_0 \int^{\infty}_{0}\,dp \nonumber \\
  && \mbox{ ln }\left\{\left[\omega_f^2 + M_{f}^2(\omega_f^2)\right]\,\frac{1}{\omega_f^2
      + m_{f}^2}\right\} \nonumber \\
  && -\,\frac{N_c}{\pi^2}\, \sum_{f=u,d,s}\,\int^{\sqrt{ \widetilde{\mu}_f^2-m_{f}^2}}_{0} \,
  dp\,p^2 \,
  \left[\,(\widetilde{\mu}_f-E_f)\, \theta(\widetilde{\mu}_f-m_f)\,\right]\nonumber \\
  &&  -\; \frac{1}{2}\left[ \sum_{f=u,d,s} (\bar \sigma_f \ \bar S_f  +
    \frac{G_s}{2} \ \bar S_f^2) \; + \; \frac{H}{2} \, \bar S_u\ \bar S_d\ \bar S_s
  \right]\nonumber \\
  && -\;\sum_{f=u,d,s} \frac{{ \varpi^2_{V,f}}}{4 G_V}, \label{omzerot}
\end{eqnarray}
where $N_c=3$, $E_{f}=\sqrt{\vec{p}\,^{2}+m_{f}^{2}}$ and we have defined
\begin{equation}
\omega_f^2 = (\,p_0\, + \,i\, \mu_f\,)^2\, + \,\vec{p}\,^2.
\end{equation}
The masses of free quarks are denoted by $m_f$, where $f=u,d,s$.  The
$\varpi_{V,f}$ mean field is related to the vector current density
$j_{V,f}^{\mu}$ of Eq.\ (\ref{currents}).

The momentum-dependent constituent quark masses $M_{f}$ depend
explicitly on the quark mean fields $\bar\sigma_{f}$,
\begin{equation}
  M_{f}(\omega_{f}^2) \ = \ m_f\, + \, \bar\sigma_f\,
  g(\omega_{f}^2), \label{qmass}
\end{equation}
where $g(\omega^2)$ denotes the Fourier transform of the form factor
$\widetilde{g}(z)$.

Following Ref.\ \cite{Weise2011}, the quark vector interaction shifts the
quark chemical potential according to
\begin {equation}
\widetilde{\mu}_f = \mu_f -  \varpi_{V,f} .
\label{eq:mu_f}
\end{equation}
Note that the shifting of the quark chemical potential does not affect
the nonlocal form factor $g(\omega_{f}^2)$ as discussed in
\cite{Weise2011}.  The mean-field values of the auxiliary fields
$\bar S_f$ are given by \citep{Scarpettini2004}
\begin{equation}
  \bar S_f = -\, 16\,N_c\, \int^{\infty}_{0}\,dp_0 \int^{\infty}_{0}
  \frac{dp}{(2\pi)^3}\,g(\omega_f^2)\,\frac{
    M_{f}(\omega_f^2)}{\omega_f^2 + M_{f}^2(\omega_f^2)}\;.
\end{equation}
In this paper we adopt a Gaussian form for the nonlocal form factor $g$, 
\begin{equation}
  g(\omega^2) = \exp{\left(-\omega^2/\Lambda^2\right)}\, ,
\end{equation}
where $\Lambda$ plays a role for the stiffness of the chiral
transition. This parameter, together with the current quark mass
$\bar{m}$ of up and down quarks and the coupling constants $G_s$
and $H$ in Eq.\ (\ref{omzerot}), have been fitted to the pion decay
constant, $f_\pi$, and meson masses $m_{\pi}$, $m_\eta$, and
$m_{\eta'}$, as described in \citep{Contrera2008,Contrera2010}. The
result of this fit is $\bar{m} = 6.2$~MeV, $\Lambda = 706.0$~MeV,
$G_s \Lambda^2 = 15.04$, $H \Lambda^5 = - 337.71$. The strange quark
current mass is treated as a free parameter and was set to $m_s
=140.7$~MeV.  The strength of the vector interaction $G_V$ is usually
expressed in terms of the strong coupling constant $G_s$. To account
for the uncertainty in the theoretical predictions for the ratio
$G_V/G_s$, we treat the vector coupling constant as a free parameter
\cite{Sasaki:2006ws,Fukushima:2008wg,Bratovic:2012qs}, which varies
from $0$ to $0.1 \, G_s$.

Using these parametrizations, the fields $\bar \sigma_f$ and
$\varpi_{V,f}$ can be determined by minimizing Eq.\ (\ref{omzerot}),
\begin{equation}
\label{nonloq}
\frac{\partial \Omega^{NL}}{\partial \bar \sigma_f} =
\frac{\partial \Omega^{NL}}{\partial {\varpi_{V,f}}}= 0 .
\end{equation}

\vskip 0.5cm
\noindent {\it Description of confined hadronic matter --} The
hadronic phase is described in the framework of the non-linear
relativistic mean field theory \cite{Walecka1974,
  Serot1986,glendenning00:book,weber99:book}, where baryons (neutrons,
protons, hyperons) interact via the exchange of scalar, vector and
isovector mesons ($\sigma$, $\omega $, $\vec \rho $, respectively).
The Lagrangian of the theory is given by
\begin{eqnarray}
  \mathcal{L} &=& \sum_{B=n,p, \Lambda, \Sigma, \Xi}\bar{\psi}_B
  \bigl[\gamma_\mu(i\partial^\mu-g_\omega
  \omega^\mu-g_\rho \vec{\rho}^\mu) \nonumber\\
  &-&(m_N-g_\sigma\sigma)\bigr]\psi_B+\frac{1}{2}(\partial_\mu\sigma\partial^\mu
  \sigma-m_\sigma^2\sigma^2) \nonumber\\
  &-&\frac{1}{3}b_\sigma m_N(g_\sigma\sigma)^3 - \frac{1}{4}c_\sigma(g_\sigma\sigma)^4-
  \frac{1}{4}\omega_{\mu\nu}\omega^{\mu\nu} \nonumber\\
  &+&\frac{1}{2} m_\omega^2 \, \omega_\mu\omega^\mu + \frac{1}{2} m_\rho^2\, 
  \vec{\rho}_\mu\cdot  \vec{\rho\,}^\mu \label{eq:lag}\\
  &-&\frac{1}{4} \vec{\rho}_{\mu\nu} \vec{\rho\,}^{\mu\nu} + \sum_{\lambda=e^-, \mu^-}
  \bar{\psi}_\lambda  (i\gamma_\mu\partial^\mu-m_\lambda)\psi_\lambda \, ,
  \nonumber
\end{eqnarray}
where $B$ sums all baryon states which become populated in neutron
star matter \cite{glendenning00:book,weber99:book}. The quantities
$g_\rho$, $g_\sigma$, and $g_\omega$ are the meson-baryon coupling
constants. Non-linear $\sigma$-meson self-interactions are taken into
account in Eq.\ (\ref{eq:lag}) via the terms proportional to
$b_\sigma$ and $c_\sigma$ \cite{glendenning00:book,weber99:book}. We
have solved the equations of motion for the baryon and meson field
equations, which follow from Eq.\ (\ref{eq:lag}), for the relativistic
mean-field approximation \citep{glendenning00:book,weber99:book}. For
this approximation the meson fields $\sigma$, $\omega$, $\rho$ are
approximated by their respective mean-field values $\bar{\sigma}
\equiv \langle\sigma\rangle$, $\bar\omega \equiv
\langle\omega\rangle$, and $\bar{\rho} \equiv \langle\rho_{03}\rangle$
\cite{glendenning00:book,weber99:book}.  The parameters of the model,
labeled GM1, are adjusted to the properties of nuclear matter at
saturation density. They are taken from \citep{Glendenning1991}.

The condition of weak equilibrium requires the presence of electrons
\begin{figure}[htb]
\includegraphics[width=0.50 \textwidth]{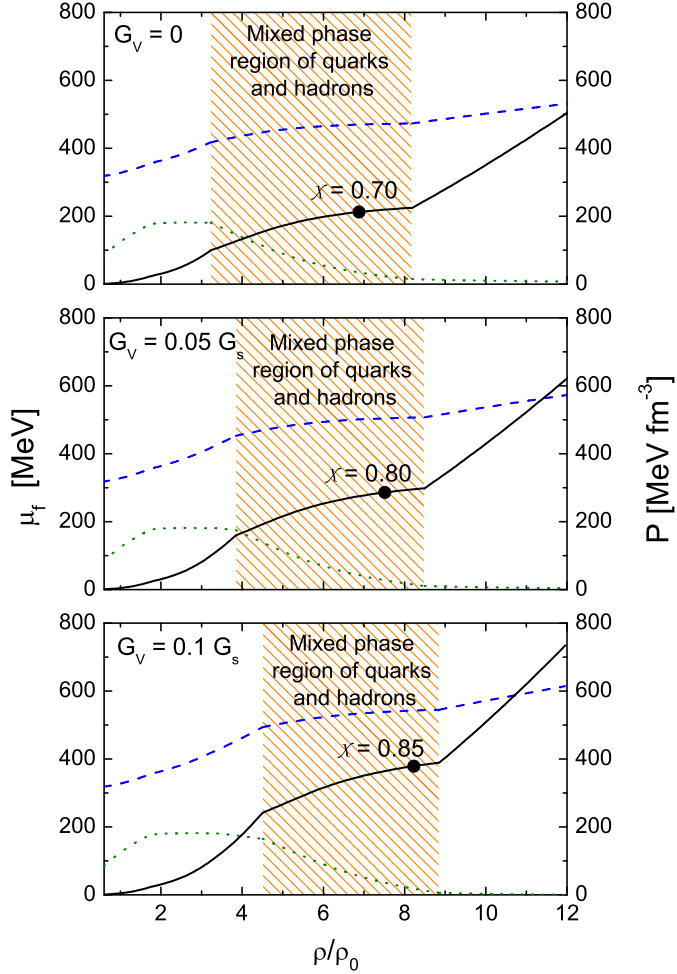}
\caption{(Color online) Pressure, $P$ (solid lines), baryon chemical
  potential, $\mu_b$ (dashed lines), and electron chemical potential,
  $\mu_e$ (dotted lines) as a function of baryon number density,
  $\rho$, in units of the normal nuclear matter density, $\rho_0 =
  0.16~{\rm fm}^{-3}$. The hatched areas denote the mixed phase
  regions where confined hadronic matter and deconfined quark matter
  coexist. The solid dots indicate the central densities of the
  associated maximum-mass stars, shown in Fig.\ \ref{masrad},
    and $\chi$ is the respective fraction of quark matter inside of
    them. The results are computed for three different values of the
  vector coupling constant, ranging from 0, to 0.05 $G_s$, to 0.1
  $G_s$.}\label{press}
\end{figure}
and muons, which are treated as free relativistic quantum gases, as
described by the last term on the right-hand-side of Eq.\
(\ref{eq:lag}). Neutron star matter is characterized by the
conservation of electric and baryon number. This feature leads to the
chemical equilibrium condition
\begin{eqnarray}
\mu_i = B_i \, \mu_n - Q_i \, \mu_e \, ,
\label{eq:mu_i}
\end{eqnarray}
where $\mu_n$ and $\mu_e$ denote the chemical potentials of neutrons
and electrons, respectively.
\begin{figure}[htb]
\includegraphics[width=0.50 \textwidth]{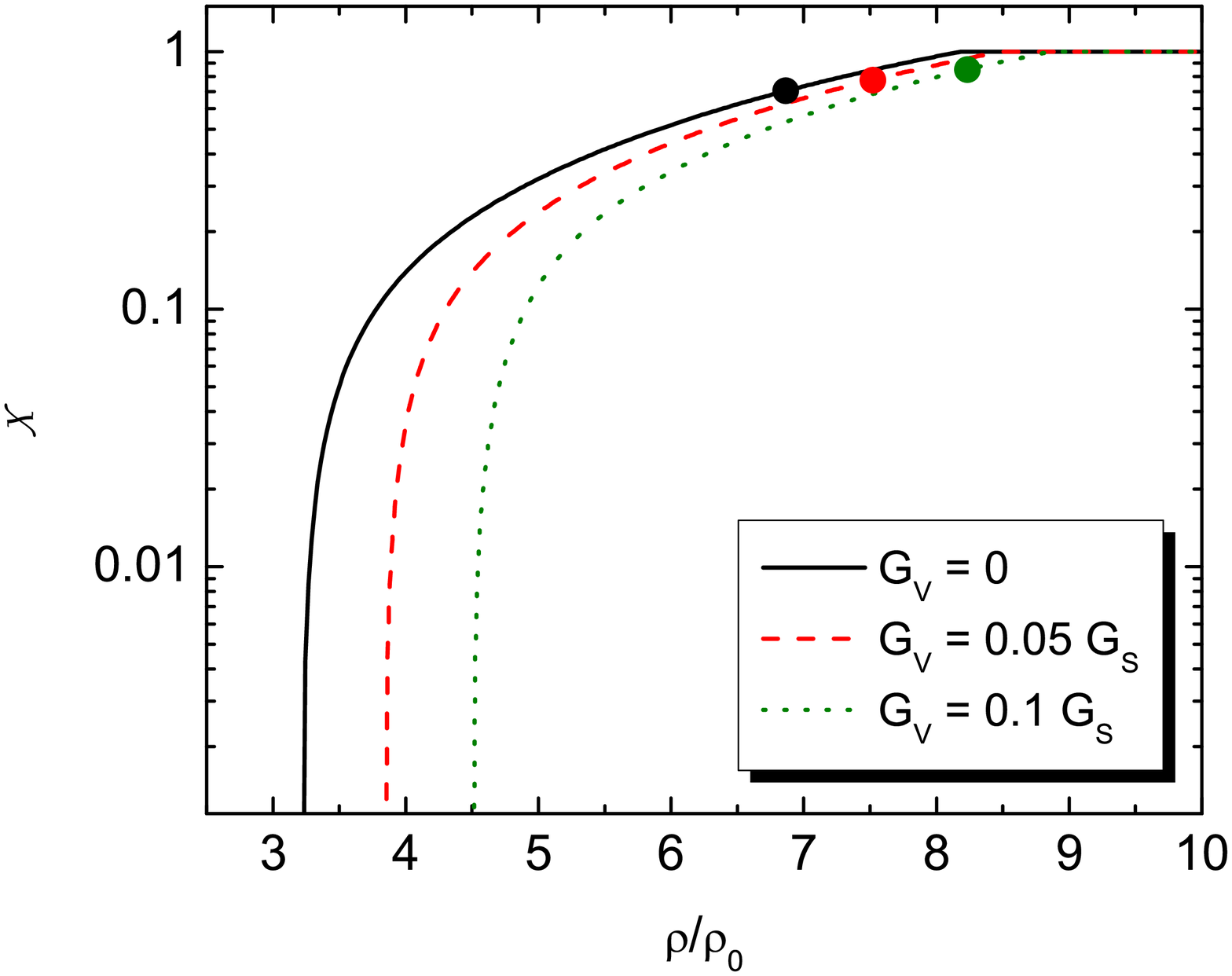}
\caption{(Color online) Volume fraction, $\chi$, of quark phase as a
  function of baryon number density, $\rho$, in units of normal
  nuclear matter density, $\rho_0 = 0.16~{\rm fm}^{-3}$. The solid
  dots indicate the central densities of the respective maximum-mass
  stars shown in Fig.\ \ref{masrad}.}\label{chiplot}
\end{figure}
The quantities $B_i$ and $Q_i$ stand for
the baryon numbers and the electric charges of the mesons and baryons
of Eq.\ (\ref{eq:lag}).  Equation (\ref{eq:mu_i}) greatly simplifies
the mathematical analysis, since only the knowledge of two independent
chemical potentials, $\mu_n$ and $\mu_e$, is necessary. The latter are
obtained from \cite{glendenning00:book,weber99:book}
\begin{align}
  \mu_B &= g_{\omega} \bar{\omega} + g_{\rho} \bar{\rho_{03}}
  I^3_B + \sqrt{k_B^2+m_B^{*2}} \,,
  \nonumber \\
  \mu_\lambda &= \sqrt{k_\lambda^2+m_\lambda^2} \, ,
\end{align}
where $m_B^*=m_B - g_\sigma \bar{\sigma}$ denote the effective
medium-modified baryon masses, $k_B$ and $k_\lambda$ are the Fermi
momenta of baryons and leptons, respectively, and $I^3_B$ is the third
component of the isospin vector of a baryon of type $B$. Finally,
aside from chemical equilibrium, the condition of electric charge
neutrality for confined hadronic matter is also of critical importance
for the composition of neutron star matter. This condition is given by
\cite{glendenning00:book,weber99:book}
\begin{eqnarray}
  \sum_B Q_i \, (2J_B+1) \, \frac{k_B^3}{6 \pi^2} -
  \sum_\lambda \frac{k_\lambda^3}{3 \pi^2} = 0 \, ,
\label{eq:electric}
\end{eqnarray}
where $J_B$ denotes the spin of baryon $B$.

For the quark phase, the chemical potentials associated with quarks
and electrons follow from Eq.\ (\ref{eq:mu_i}) as $\mu_u = \mu_b - 2
\mu_e /3$ and $\mu_d = \mu_s = \mu_b + \mu_e /3$, where $\mu_b = \mu_n
/ 3$ stands for the baryon chemical potential, expressed in terms of
the chemical potential of neutrons.

\vskip 0.5cm
\noindent {\it Description of the mixed phase of quarks and
  hadrons --} To determine the mixed phase region of quarks and hadrons, we start
from the Gibbs condition for phase equilibrium between hadronic ($H$) and
quark ($Q$) matter,
\begin{eqnarray}
  P_H ( \mu_n , \mu_e, \{ \phi \} ) = P_Q (\mu_n , \mu_e) \, ,
\label{eq:gibbs}
\end{eqnarray}
where $P_H$ and $P_Q$ denote the pressures of hadronic matter and
quark matter, respectively \cite{glendenning92:a}. The quantity $\{
\phi \}$ in Eq.\ (\ref{eq:gibbs}) stands collectively for the field
variables ($\bar{\sigma}$, $\bar\omega$, $\bar\rho$) and Fermi momenta
($k_B$, $k_\lambda$) that characterize a solution to the equations of
confined hadronic matter. We use the symbol $\chi \equiv V_Q/V$ to
denote the volume proportion of quark matter, $V_Q$, in the unknown
volume $V$. By definition, $\chi$ then varies between 0 and 1,
depending on how much confined hadronic matter has been converted to
quark matter.  Equation (\ref{eq:gibbs}) is to be supplemented with
the conditions of global baryon charge conservation and global
electric charge conservation.  The global conservation of baryon
charge is expressed as \cite{glendenning92:a}
\begin{eqnarray}
  \rho_b = \chi \, \rho_Q(\mu_n, \mu_e ) + (1-\chi) \,
  \rho_H (\mu_n, \mu_e,  \{ \phi \}) \, ,
\label{eq:mixed_rho}
\end{eqnarray}
where  $\rho_Q$ and $\rho_H$ denote the baryon number densities of the
quark phase and hadronic phase, respectively. The global neutrality of
electric charge is given by \cite{glendenning92:a}
\begin{align}
  0 = \chi \ q_Q(\mu_n, \mu_e ) + (1-\chi) \ q_H (\mu_n, \mu_e, \{
  \phi \}) \, ,
\label{eq:mixed_charge}
\end{align}
with $q_Q$ and $q_H$ denoting the electric charge densities of the
quark phase and hadronic phase, respectively.  We have chosen global
rather than local electric charge neutrality, since the latter is not
fully consistent with the Einstein-Maxwell equations and the micro
physical condition of $\beta$--equilibrium and relativistic quantum
statistics, as shown in \citep{Rufini2011}. Local NJL studies carried
out for local electric charge neutrality have been reported recently
in Refs.\ \cite{Masuda2012,Lenzi2012}. In
  Ref. \cite{Lenzi2012} the nonlinear relativistic mean-field model
  GM1 and the local SU(3) NJL model with vector interaction were used
  to describe the hadronic and quark matter phases, respectively.  The
  authors found that the observation of neutron stars with masses a
  few percent higher than the $1.97 \pm 0.04 \, M_{\odot}$ would be
  hard to explain unless, instead of GM1, one uses a very stiff model
  for the hadronic EOS with nucleons only. They obtain neutron stars
  with stable pure quark matter cores in their centers. In contrast to
  this, we find that such neutron stars are not be stable if the
  nonlocal NJL model is used instead of the local model and the less
  stringent condition of global electric charge neutrality is
  imposed on the composition of the stellar matter.

\begin{figure}
     \begin{center}
       \subfigure[ Mass-radius relationships of neutron stars with
       quark-hybrid matter in their centers.]{%
            \label{fig:first}
            \includegraphics[width=0.450\textwidth]{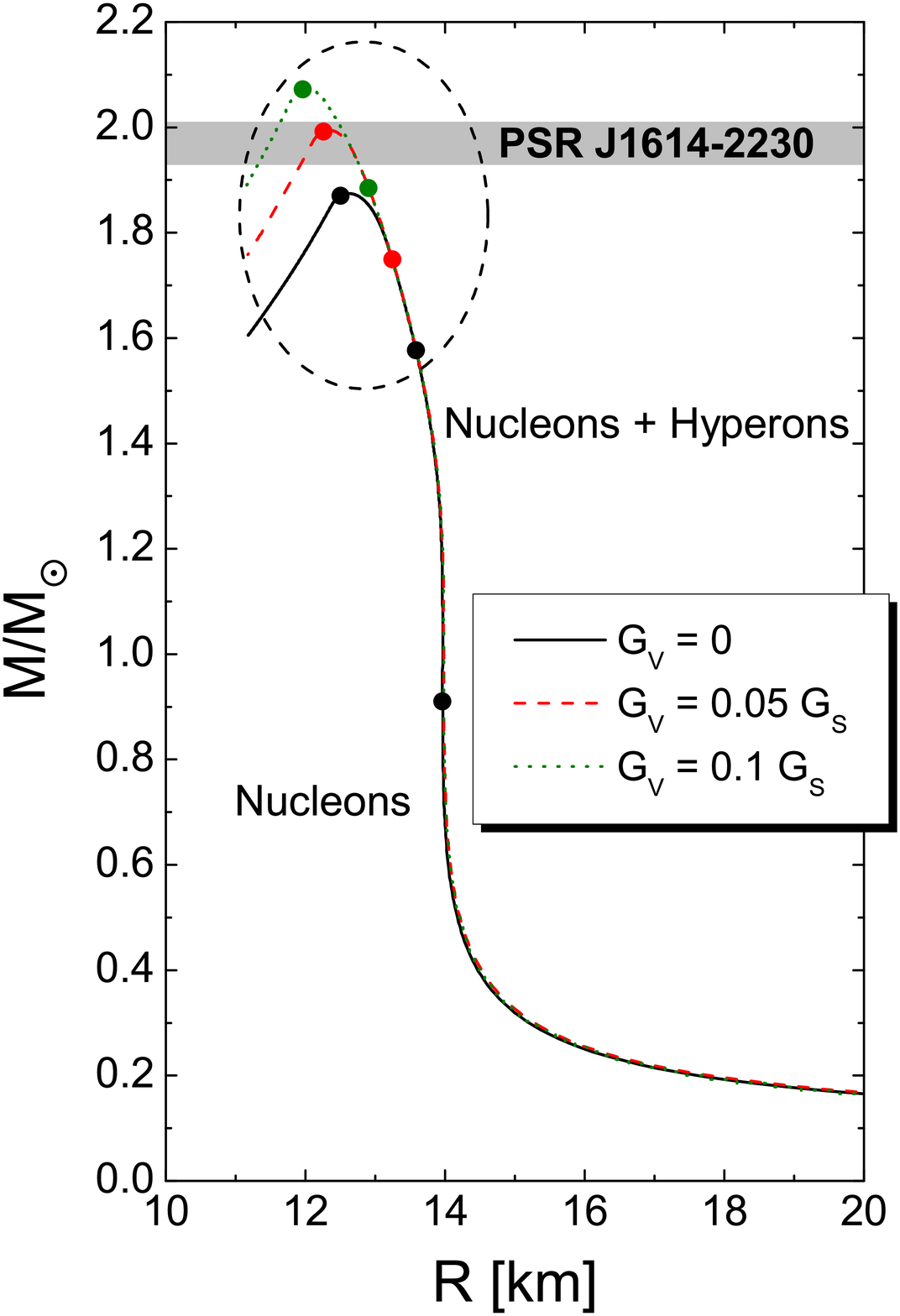}}
          \subfigure[ Enlargement of the circled region of Fig.\
          \ref{fig:first}.  The labels 'MP' and 'QP' stand for mixed
          phase and pure quark phase, respectively. The solid vertical
          bars denote maximum-mass stars.]{%
           \label{fig:second}
           \includegraphics[width=0.45\textwidth]{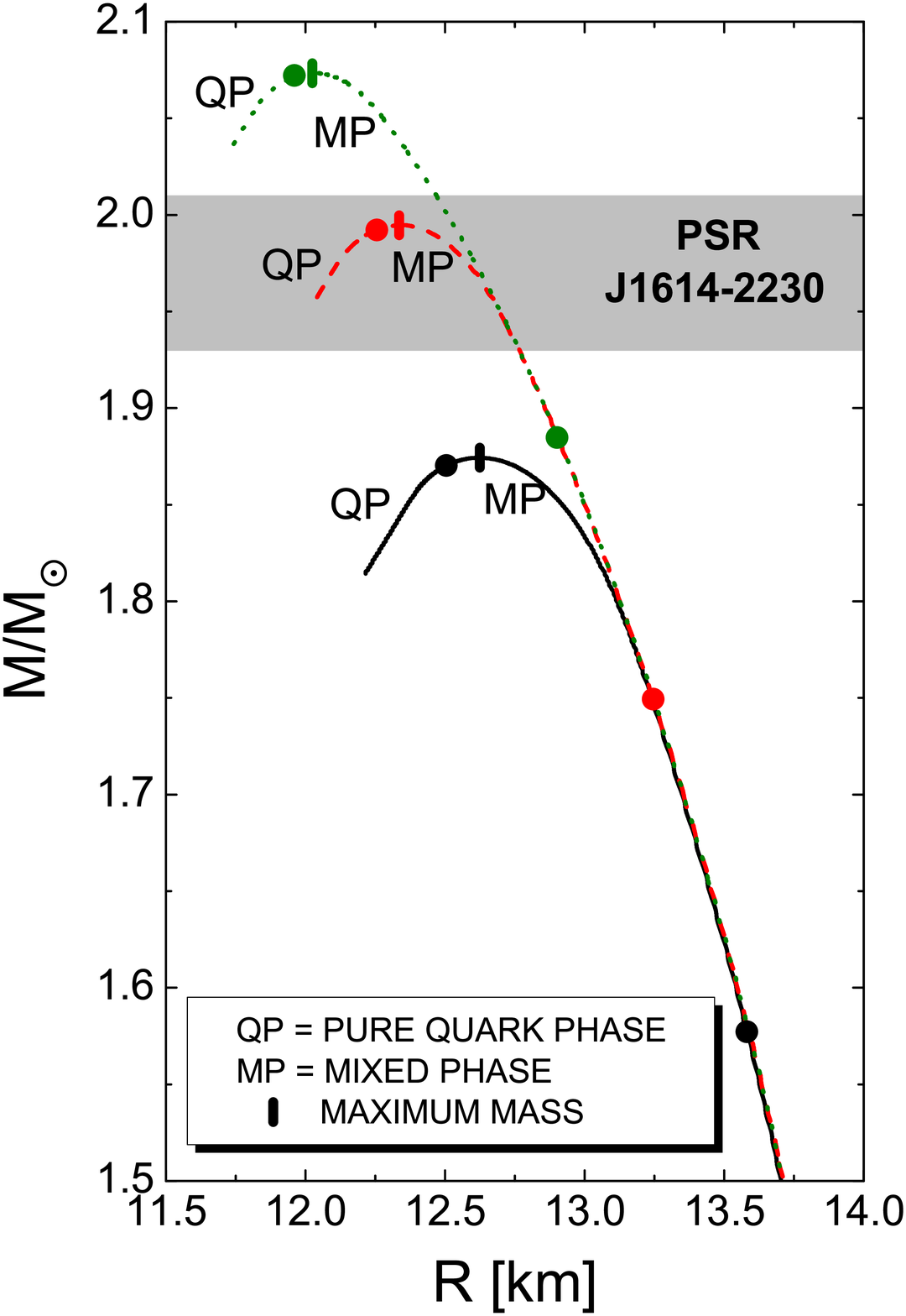}}
    \end{center}
    \caption{(Color online) Depending on the strength of the vector
      repulsion ($G_V$) of the nonlocal NJL model, maximum masses up to $2.1\,
      M_\odot$ are obtained.  With increasing stellar mass, the stellar
      core compositions consist of either only nucleons, nucleons and
      hyperons, a mixed phase of quarks and hadrons (MP), or a pure
      quark matter phase (QP). The latter, however, exists
      only in neutron stars which lie to the left of their respective
      mass peaks. Such stars are unstable against radial
      oscillations and thus cannot exist stably in the universe. In
      contrast to this, all neutron stars on the MP branches up to
      the mass peaks are stable.  The gray band denotes the 1-$\sigma$
      error bar of the $M = (1.97 \pm 0.04)M_\odot$ neutron star PSR
      J1614-2230 \citep{Demorest2010}.}
   \label{masrad}
\end{figure}

The results for the mixed phase region are shown in Figs.\
  \ref{press} and \ref{chiplot}. Our calculations show that the
inclusion of the quark vector coupling contribution shifts the onset
of the phase transition to higher densities, and also narrows the
width of the mixed quark-hadron phase, when compared to the case $G_V
= 0$. The mixed phases range from $3.2 - 8.2 \rho_0$, $3.8 - 8.5
\rho_0$, and $4.5 - 8.9 \rho_0$ for vector coupling constants $G_V/G_s
= 0, \ 0.05 , \ 0.1$, respectively. We note that there is
  considerable theoretical uncertainty in the ratio of $G_{V}/G_{s}$
  \cite{Dumm} since a rigorous derivation of the effective couplings
  from QCD is not possible. Combining the ratios of $G_{V}/G_{s}$ from
  the molecular instanton liquid model and from the Fierz
  transformation, the value of $G_{V}/G_{s}$ is expected to be in the
  range $0 \leqslant G_{V}/G_{s} \leqslant 0.5$ \cite{Zhang}. For our
  model, values of $G_{V}/G_{s} > 0.1$ shift the onset of the
  quark-hadron phase transition to such high densities that quark
  deconfinement can not longer occur in the cores of neutron stars.

Next we determine the bulk properties of spherically symmetric neutron
stars for the collection of equations of state shown in Fig.\
\ref{press}. The properties are determined by solving the
Tolmann-Oppenheimer-Volkoff (TOV) equation of general relativity
theory \citep{Tolman39}.  The outcome is shown in Fig.\ \ref{masrad}.
One sees that depending on the value of the vector coupling constant,
$G_V$, the maximum neutron star masses increase from $1.87 \, M_\odot$
for $G_V = 0$, to $2.00 \, M_\odot$ for $G_V = 0.05 \, G_s$, to $2.07
\,M_\odot$ for $G_V = 0.1 \, G_s$. The heavier stars of all three
stellar sequences contain mixed phases of quarks and hadrons in their
centers. The densities in these stars are however not high enough to
generate pure quark matter in the cores. Such matter forms only in
neutron stars which are already located on the gravitationally
unstable branch of the mass-radius relationships. Another intriguing 
finding is that neutron stars with canonical masses of around 1.4 $M_\odot$ 
do not posses a mixed phase of quarks and hadrons but are made entirely of 
confined hadronic matter.

\vskip 0.5cm
\noindent{\it Summary and Conclusions --} In this paper, we show that
high-mass neutron star, such as PSR J1614--2230 with a gravitational
mass of $1.97\pm 0.04\, M_{\odot}$ \citep{Demorest2010}, may contain
mixtures of quarks and hadrons in their central regions.  Our analysis
is based on a nonlocal extension of the SU(3) Nambu-Jona Lasinio
model, which reproduces some of the key features of Quantum
Chromodynamics at densities relevant to neutron stars. Critical is the
inclusion of the quark vector coupling contribution in the nonlocal
SU(3) NJL model. Our results also show that the transition to pure
quark matter occurs only in neutron stars which lie already on the
gravitationally unstable branch of the mass-radius relationship. The
existence of pure quark matter in massive (as well as in all other,
less massive) neutron stars would thus be ruled out by our study.

\acknowledgments 

\noindent
M.\ Orsaria thanks N.\ N.\ Scoccola for fruitful discussions about the
nonlocal NJL model.  M.\ Orsaria and G.\ Contrera thank CONICET for
financial support. H.\ Rodrigues thanks CAPES for financial support
under contract number BEX 6379/10-9. F.\ Weber is supported by the
National Science Foundation (USA) under Grant PHY-0854699.

\end{document}